\documentclass[%
 reprint, 
nofootinbib,
 amsmath,amssymb,
 aps,
 prl,superscriptaddress
]{revtex4-1}

\usepackage{tabularx}

\usepackage{graphicx}
\usepackage{dcolumn}
\usepackage{bm}
\usepackage{hyperref}
\usepackage[all]{hypcap} 
\usepackage{tabularx}

\usepackage[T5,T1]{fontenc}
\usepackage{placeins}

\newcommand{\eab}{\epsilon_{\alpha \beta}}

\newcommand{\eut}{\epsilon_{\mu \tau}}
\newcommand{\ett}{\epsilon_{\tau \tau}}

\newcommand{\mnu}{\nu_{\mu}}
\newcommand{\onu}{\overline{\nu}_{\mu}}
\newcommand{\nt}{\nu_{\tau}}
\newcommand{\ot}{\overline{\nu}_{\tau}}

\begin{document}

\preprint{APS/123-QED}
\setlength{\textfloatsep}{12pt}
\setlength{\abovecaptionskip}{0pt}
\setlength{\belowcaptionskip}{12pt}


\title{Strong constraints on neutrino nonstandard interactions from TeV-scale \texorpdfstring{$\nu_\mu$}{nm} disappearance at IceCube}

\affiliation{III. Physikalisches Institut, RWTH Aachen University, D-52056 Aachen, Germany}
\affiliation{Department of Physics, University of Adelaide, Adelaide, 5005, Australia}
\affiliation{Dept. of Physics and Astronomy, University of Alaska Anchorage, 3211 Providence Dr., Anchorage, AK 99508, USA}
\affiliation{Dept. of Physics, University of Texas at Arlington, 502 Yates St., Science Hall Rm 108, Box 19059, Arlington, TX 76019, USA}
\affiliation{CTSPS, Clark-Atlanta University, Atlanta, GA 30314, USA}
\affiliation{School of Physics and Center for Relativistic Astrophysics, Georgia Institute of Technology, Atlanta, GA 30332, USA}
\affiliation{Dept. of Physics, Southern University, Baton Rouge, LA 70813, USA}
\affiliation{Dept. of Physics, University of California, Berkeley, CA 94720, USA}
\affiliation{Lawrence Berkeley National Laboratory, Berkeley, CA 94720, USA}
\affiliation{Institut f{\"u}r Physik, Humboldt-Universit{\"a}t zu Berlin, D-12489 Berlin, Germany}
\affiliation{Fakult{\"a}t f{\"u}r Physik {\&} Astronomie, Ruhr-Universit{\"a}t Bochum, D-44780 Bochum, Germany}
\affiliation{Universit{\'e} Libre de Bruxelles, Science Faculty CP230, B-1050 Brussels, Belgium}
\affiliation{Vrije Universiteit Brussel (VUB), Dienst ELEM, B-1050 Brussels, Belgium}
\affiliation{Department of Physics and Laboratory for Particle Physics and Cosmology, Harvard University, Cambridge, MA 02138, USA}
\affiliation{Dept. of Physics, Massachusetts Institute of Technology, Cambridge, MA 02139, USA}
\affiliation{Dept. of Physics and The International Center for Hadron Astrophysics, Chiba University, Chiba 263-8522, Japan}
\affiliation{Department of Physics, Loyola University Chicago, Chicago, IL 60660, USA}
\affiliation{Dept. of Physics and Astronomy, University of Canterbury, Private Bag 4800, Christchurch, New Zealand}
\affiliation{Dept. of Physics, University of Maryland, College Park, MD 20742, USA}
\affiliation{Dept. of Astronomy, Ohio State University, Columbus, OH 43210, USA}
\affiliation{Dept. of Physics and Center for Cosmology and Astro-Particle Physics, Ohio State University, Columbus, OH 43210, USA}
\affiliation{Niels Bohr Institute, University of Copenhagen, DK-2100 Copenhagen, Denmark}
\affiliation{Dept. of Physics, TU Dortmund University, D-44221 Dortmund, Germany}
\affiliation{Dept. of Physics and Astronomy, Michigan State University, East Lansing, MI 48824, USA}
\affiliation{Dept. of Physics, University of Alberta, Edmonton, Alberta, Canada T6G 2E1}
\affiliation{Erlangen Centre for Astroparticle Physics, Friedrich-Alexander-Universit{\"a}t Erlangen-N{\"u}rnberg, D-91058 Erlangen, Germany}
\affiliation{Physik-department, Technische Universit{\"a}t M{\"u}nchen, D-85748 Garching, Germany}
\affiliation{D{\'e}partement de physique nucl{\'e}aire et corpusculaire, Universit{\'e} de Gen{\`e}ve, CH-1211 Gen{\`e}ve, Switzerland}
\affiliation{Dept. of Physics and Astronomy, University of Gent, B-9000 Gent, Belgium}
\affiliation{Dept. of Physics and Astronomy, University of California, Irvine, CA 92697, USA}
\affiliation{Karlsruhe Institute of Technology, Institute for Astroparticle Physics, D-76021 Karlsruhe, Germany }
\affiliation{Karlsruhe Institute of Technology, Institute of Experimental Particle Physics, D-76021 Karlsruhe, Germany }
\affiliation{Dept. of Physics, Engineering Physics, and Astronomy, Queen's University, Kingston, ON K7L 3N6, Canada}
\affiliation{Dept. of Physics and Astronomy, University of Kansas, Lawrence, KS 66045, USA}
\affiliation{Department of Physics and Astronomy, UCLA, Los Angeles, CA 90095, USA}
\affiliation{Centre for Cosmology, Particle Physics and Phenomenology - CP3, Universit{\'e} catholique de Louvain, Louvain-la-Neuve, Belgium}
\affiliation{Department of Physics, Mercer University, Macon, GA 31207-0001, USA}
\affiliation{Dept. of Astronomy, University of Wisconsin{\textendash}Madison, Madison, WI 53706, USA}
\affiliation{Dept. of Physics and Wisconsin IceCube Particle Astrophysics Center, University of Wisconsin{\textendash}Madison, Madison, WI 53706, USA}
\affiliation{Institute of Physics, University of Mainz, Staudinger Weg 7, D-55099 Mainz, Germany}
\affiliation{Department of Physics, Marquette University, Milwaukee, WI, 53201, USA}
\affiliation{Institut f{\"u}r Kernphysik, Westf{\"a}lische Wilhelms-Universit{\"a}t M{\"u}nster, D-48149 M{\"u}nster, Germany}
\affiliation{Bartol Research Institute and Dept. of Physics and Astronomy, University of Delaware, Newark, DE 19716, USA}
\affiliation{Dept. of Physics, Yale University, New Haven, CT 06520, USA}
\affiliation{Dept. of Physics, University of Oxford, Parks Road, Oxford OX1 3PU, UK}
\affiliation{Dept. of Physics, Drexel University, 3141 Chestnut Street, Philadelphia, PA 19104, USA}
\affiliation{Physics Department, South Dakota School of Mines and Technology, Rapid City, SD 57701, USA}
\affiliation{Dept. of Physics, University of Wisconsin, River Falls, WI 54022, USA}
\affiliation{Dept. of Physics and Astronomy, University of Rochester, Rochester, NY 14627, USA}
\affiliation{Department of Physics and Astronomy, University of Utah, Salt Lake City, UT 84112, USA}
\affiliation{Oskar Klein Centre and Dept. of Physics, Stockholm University, SE-10691 Stockholm, Sweden}
\affiliation{Dept. of Physics and Astronomy, Stony Brook University, Stony Brook, NY 11794-3800, USA}
\affiliation{Dept. of Physics, Sungkyunkwan University, Suwon 16419, Korea}
\affiliation{Institute of Basic Science, Sungkyunkwan University, Suwon 16419, Korea}
\affiliation{Institute of Physics, Academia Sinica, Taipei, 11529, Taiwan}
\affiliation{Dept. of Physics and Astronomy, University of Alabama, Tuscaloosa, AL 35487, USA}
\affiliation{Dept. of Astronomy and Astrophysics, Pennsylvania State University, University Park, PA 16802, USA}
\affiliation{Dept. of Physics, Pennsylvania State University, University Park, PA 16802, USA}
\affiliation{Dept. of Physics and Astronomy, Uppsala University, Box 516, S-75120 Uppsala, Sweden}
\affiliation{Dept. of Physics, University of Wuppertal, D-42119 Wuppertal, Germany}
\affiliation{DESY, D-15738 Zeuthen, Germany}

\author{R. Abbasi}
\affiliation{Department of Physics, Loyola University Chicago, Chicago, IL 60660, USA}
\author{M. Ackermann}
\affiliation{DESY, D-15738 Zeuthen, Germany}
\author{J. Adams}
\affiliation{Dept. of Physics and Astronomy, University of Canterbury, Private Bag 4800, Christchurch, New Zealand}
\author{J. A. Aguilar}
\affiliation{Universit{\'e} Libre de Bruxelles, Science Faculty CP230, B-1050 Brussels, Belgium}
\author{M. Ahlers}
\affiliation{Niels Bohr Institute, University of Copenhagen, DK-2100 Copenhagen, Denmark}
\author{M. Ahrens}
\affiliation{Oskar Klein Centre and Dept. of Physics, Stockholm University, SE-10691 Stockholm, Sweden}
\author{J.M. Alameddine}
\affiliation{Dept. of Physics, TU Dortmund University, D-44221 Dortmund, Germany}
\author{C. Alispach}
\affiliation{D{\'e}partement de physique nucl{\'e}aire et corpusculaire, Universit{\'e} de Gen{\`e}ve, CH-1211 Gen{\`e}ve, Switzerland}
\author{A. A. Alves Jr.}
\affiliation{Karlsruhe Institute of Technology, Institute for Astroparticle Physics, D-76021 Karlsruhe, Germany }
\author{N. M. Amin}
\affiliation{Bartol Research Institute and Dept. of Physics and Astronomy, University of Delaware, Newark, DE 19716, USA}
% ...
\date{\today}

\collaboration{IceCube Collaboration}
\noaffiliation

\begin{abstract}
We report a search for nonstandard neutrino interactions (NSI) using eight years of TeV-scale atmospheric muon neutrino data from the IceCube Neutrino Observatory.  By reconstructing incident energies and zenith angles for atmospheric neutrino events, this analysis presents unified confidence intervals for the NSI parameter $\epsilon_{\mu \tau}$. The best-fit value is consistent with no NSI at a p-value of 25.2\%. With a 90\% confidence interval of $-0.0041 \leq \epsilon_{\mu \tau} \leq 0.0031$ along the real axis and similar strength in the complex plane, this result is the strongest constraint on any NSI parameter from any oscillation channel to date.
\end{abstract}

\maketitle

\section{\label{sec:intro}Introduction}

Neutrino oscillations are a phenomenon indicating mechanisms beyond the current Standard Model (SM) of particle physics. Experiments have measured the mixing parameters of neutrino states to excellent precision and confirm that at least two states have non-zero mass~\cite{PDGreview,globalnuosc2018,deSalas2020pgw,Capozzi:2021fjo}.  Neutrino masses are orders of magnitude lighter than the other SM fermion masses, further suggesting the existence of beyond-Standard-Model (BSM) physics~\cite{numassmodels,unifiednumodels}.  

When the SM is treated as an effective field theory, neutrino masses can be introduced through the addition of a dimension-5 operator to the SM Lagrangian, with further BSM physics expected through the addition of dimension-6 operators required for renormalizability~\cite{weinbergnonconservingprocesses,wolfoscmatter,resoscmatter,numsw}.  One class of these dimension-6 operators introduces neutrino non-standard interactions (NSI), which are comprised of new neutral-current (NC) and charged-current (CC) neutrino interactions with charged fermions~\cite{nsiatmo,atmobsm,nsireactorbeam,nsibsm,cpanddegennsi}.  

This paper presents IceCube's latest constraints on the NC NSI parameter $\eut$ using eight years of muon-neutrino-\footnote{``Neutrinos'' refers to both neutrinos and antineutrinos unless otherwise stated.}induced up-going track data\footnote{The $\nu_\mu$ purity of this sample, determined from simulated neutrino and cosmic ray event simulation, is $>99.9\%$~\cite{meowsprd}.}, with the highest range of event energies (500 GeV to $\sim$10 TeV) employed for an NSI analysis to date.  A likelihood analysis is performed on the binned neutrino event counts to search for evidence of NSI via modified coherent forward scattering.  The analysis uses the same sample of neutrino events and techniques as used in the recent IceCube search for sterile neutrinos through $\nu_\mu$ disappearance, which is described in detail in Refs.~\cite{meowsprl,meowsprd}.

\begin{figure}[t]
\includegraphics[width=\columnwidth]{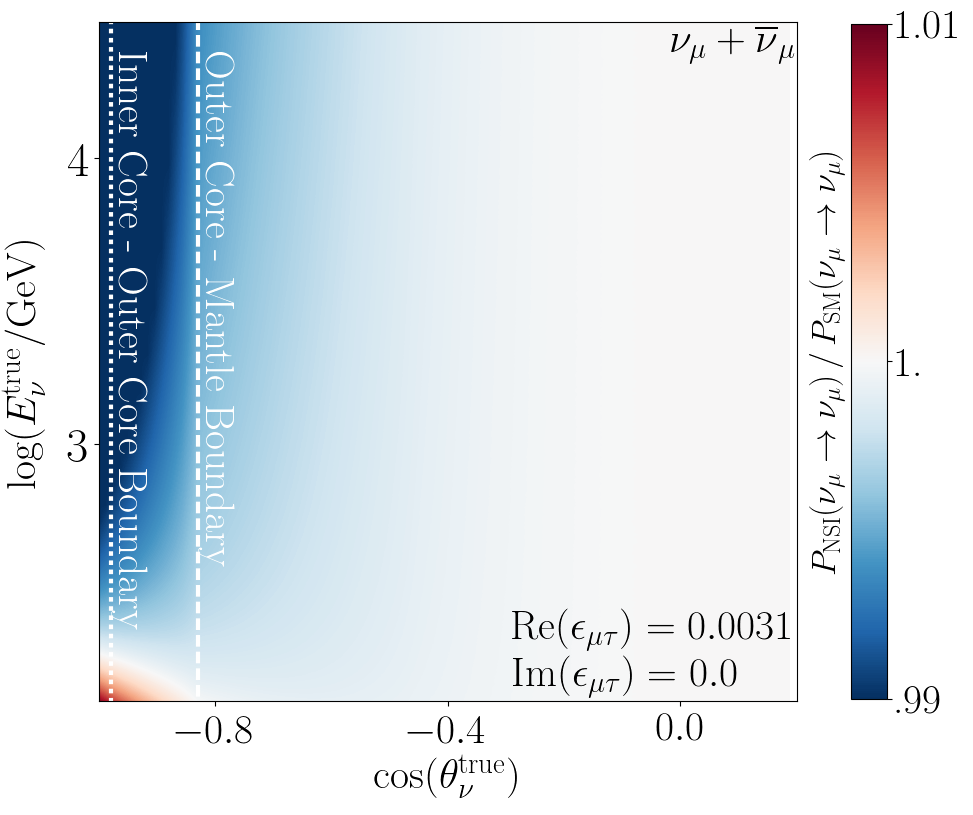}
\caption{\label{fig:epsart}\textit{\textbf{Muon neutrino oscillogram.}}  An example of how NSI modify predicted neutrino fluxes.  Shown here is the probability ratio of NSI-modified oscillations to the SM prediction for atmospheric neutrinos (chosen value is Re$(\epsilon_{\mu \tau}) = 0.0031$, Im$(\epsilon_{\mu \tau}) = 0.0$, the $90\%$ CL bound on positive Re($\eut$)).  Effects include flux disappearance at energies of 1 TeV and above for events crossing the largest Earth baselines ( $\cos (\theta) = -1)$ and flux enhancement at $\sim100$ GeV.  Note that the neutrino true energy range corresponds to the stated muon proxy energy range, and that the maximum disappearance for this value of $\epsilon_{\mu \tau}$ is $\sim 3.4\%$.}
\end{figure}


\section{\label{sec:nuosc}Neutrino Oscillations in Earth with Nonstandard Interactions}

Neutrino oscillations in matter are influenced by both material density and composition~\cite{Akhmedov,numsw,oscqft,resoscmatter,wolfoscmatter}.  For SM CC coherent scattering, the potential in the flavor basis at position $x$ is represented by~\cite{matpot},

\begin{gather}
    H_\text{mat}(x) = V_{CC}(x) \begin{pmatrix}
1 & 0 & 0\\
0 & 0 & 0\\
0 & 0 & 0\\
\end{pmatrix},
\end{gather}

\noindent with $H_{\text{mat}}\rightarrow-H_{\text{mat}}$ for antineutrinos, and the SM matter potential $V_{CC}(x) = \sqrt{2} G_F N_e(x)$, where $G_F$ is the Fermi constant and $N_e(x)$ is the electron number density~\cite{cosmocoscat,cosmonuscat}.  To include NSI from a mediator of an unknown energy scale, the collection of flavor-violating and flavor-conserving parameters $\epsilon_{\alpha \beta}$ are introduced, with indices $\alpha$ and $\beta$ corresponding to neutrino flavors \textit{$e$}, \textit{$\mu$}, and \textit{$\tau$}.  These parameters are defined through the contributions of electrons and nucleons: $ \eab \approx \eab^e + \eab^p + Y_n^\oplus \eab^n$, with $Y_n^\oplus \equiv \langle N_n(x)/N_e(x)\rangle$ where $N_e(x)$ and $N_n(x)$ are the particle number densities at matter depth $x$ for electrons and neutrons, respectively.  To good approximation, this is constant through Earth, having $Y_n^\oplus \approx 1.051$~\cite{globalnuosc2018}.  From these generalized NSI parameters, the combined matter+NSI Hamiltonian is

\begin{gather}
    H_\text{mat+NSI} = V_{CC}(x) \begin{pmatrix}
1+\epsilon_{ee} & \epsilon_{e\mu} & \epsilon_{e\tau}\\
\epsilon_{e\mu}^* & \epsilon_{\mu \mu} & \epsilon_{\mu\tau}\\
\epsilon_{e\tau}^* & \epsilon_{\mu\tau}^* & \epsilon_{\tau \tau}
\end{pmatrix},
\end{gather}

\noindent where $\epsilon^*$ is the complex conjugate of $\epsilon$ and the diagonal parameters are real-valued.  Past analyses from IceCube have set constraints on each parameter with a maximum reconstructed energy of 100 GeV~\cite{deepcore2021}. 

In this analysis, the parameter of interest is solely $\eut$, as the atmospheric neutrino flux is primarily $\mnu$ and $\onu$, which for energies $\geq20$ GeV predominantly oscillate to $\nt$ and $\ot$ due to $\nu_e$ decoupling~\cite{salvado, icecube2017}.  As a result, the atmospheric neutrino sample used in this analysis is most suitable for constraining $\mu-\tau$ flavor-changing NSI.  To verify that $\eut$ may be constrained independently, atmospheric fluxes were simulated with each NSI parameter injected at the boundary values presented by Refs.~\cite{deepcore2021,nsieuconstraint}.  Non-$\eut$ parameters, except for $\ett$, were found to induce < 0.2\%  neutrino disappearance at all sample energies and zenith angles, whereas $\eut$ = 0.0031 (analysis 90\% CL right bound) induced $\sim$3.2\% neutrino disappearance.  While for large $\eut$ the constraints on $\eut$ and $\ett$ become correlated, strong $\ett$ IceCube constraints~\cite{deepcore2021} imply the $\eut$ limit generated at $\ett$ = 0 is accurate over the allowed parameter space.  Thus, the results of this work present a standalone constraint on $\eut$. 

\section{\label{sec:nsi}\texorpdfstring{$\mu-\tau$}{mutau} NSI in IceCube}
\setlength{\parskip}{0cm}

The IceCube Neutrino Observatory is a neutrino detector located at the Geographic South Pole, occupying 1 $\text{km}^3$ of ice at depths 1450-2450 m under the Antarctic surface~\cite{icdesign}.  5160 Digital Optical Modules (DOMs)~\cite{icdaq}, each consisting of a photomultiplier tube encased in a pressurized glass sphere, are distributed in a hexagonal grid along 78 60-DOM strings spaced 125 m laterally with a vertical DOM spacing of 17 m.  An 8-string array of high quantum efficiency DOMs called DeepCore~\cite{deepcoredesign} is placed near the center of the detector at the depth where the ice is clearest.  The DeepCore string lateral spacing ranges from 42 m to 72 m, with a DOM vertical spacing of 7 m.  Data from the full array are used for event selection and reconstruction of relevant observables.

Cosmic-ray(CR)-induced air showers produce high-energy muons and neutrinos that comprise the majority of IceCube events.  While muons produced in the Southern hemisphere $(\text{``down-going'', }\cos (\theta^\text{true}_\mu) > 0)$ often penetrate the detector volume and are a background to muon neutrino signals, muons produced in the Northern hemisphere $(\cos (\theta^\text{true}_\mu) < 0)$ are absorbed by the Earth, eliminating the muon background to ``up-going'' muon neutrino signals.  A CC $\mnu$ interaction will produce hadronic products and a forward daughter muon with $\sim$ 50\%-80\% of the neutrino's energy~\cite{spencer}.  

As the muon travels it emits Cherenkov photons that are detected by IceCube DOMs, producing a track-like event that can originate either inside the detector or kilometers outside the array~\cite{astrophys,atmolep}.  From analyzing DOM charge and timing data, the zenith angle and energy of the muon are reconstructed, which determines the incident path through Earth and energy of the neutrino. This analysis uses a sample of 305,735 reconstructed muon tracks from neutrino CC interactions detected between May 13th 2011 to May 19th 2019.  Events are binned uniformly both in reconstructed muon energy $\log (E_{\text{reco}}^\mu)$ (13 bins, $E_{\text{reco}}^\mu \in [500\text{ GeV}\, ,  
9976\text{ GeV}]$) and cosine of the muon zenith angle (20 bins, $\cos(\theta_{\text{reco}}^\mu) \in [-1.0 \, , 0.0]$).  

NSI signals in IceCube manifest in the form of anomalous neutrino flavour transitions in detected events compared to the SM prediction.  When considering a neutrino-only flux (no antineutrinos) and positive values of $\text{Re}(\eut)$, there is appearance of $\nu_\mu$ due to modified $\nu_\tau$ transitions at $ E_{\nu}^{\text{true}} \lesssim 1 \text{TeV}$ and $-1 \leq \cos (\theta) \lesssim -0.8$, whereas for negative $\text{Re}(\eut)$, it is disappearance of $\nu_\mu$ in the same region.  This situation is reversed in the antineutrino case as well as in the inverted neutrino mass ordering (IO) case~\cite{inverted}.  IceCube cannot distinguish between neutrino and antineutrino signals, and thus the exact $\mnu / \onu$ sample ratio in the analysis sample is unknown\footnote{From improved hadronic models and cosmic ray measurements, the predicted ratio of atmospheric $\mnu : \onu$ is $\sim2:1$~\cite{Koshio}.}.  For equal rates of neutrinos and antineutrinos, the combined NSI effects result in NSI signals $>50\%$ weaker than what is predicted for a pure-neutrino or antineutrino sample.  An example of this is shown in Fig.~\ref{fig:epsart}, in which the NSI effect is largely energy-independent disappearance in the up-going direction.  The inability of IceCube to discriminate between neutrinos and antineutrinos also requires an independent fit to the inverted neutrino mass ordering (IO) model, which is reported in addition to the normal ordering (NO) results (Fig.~\ref{fig:1d}).

\section{Analysis}

\begin{figure}[t]
\centering
\includegraphics[width=\columnwidth]{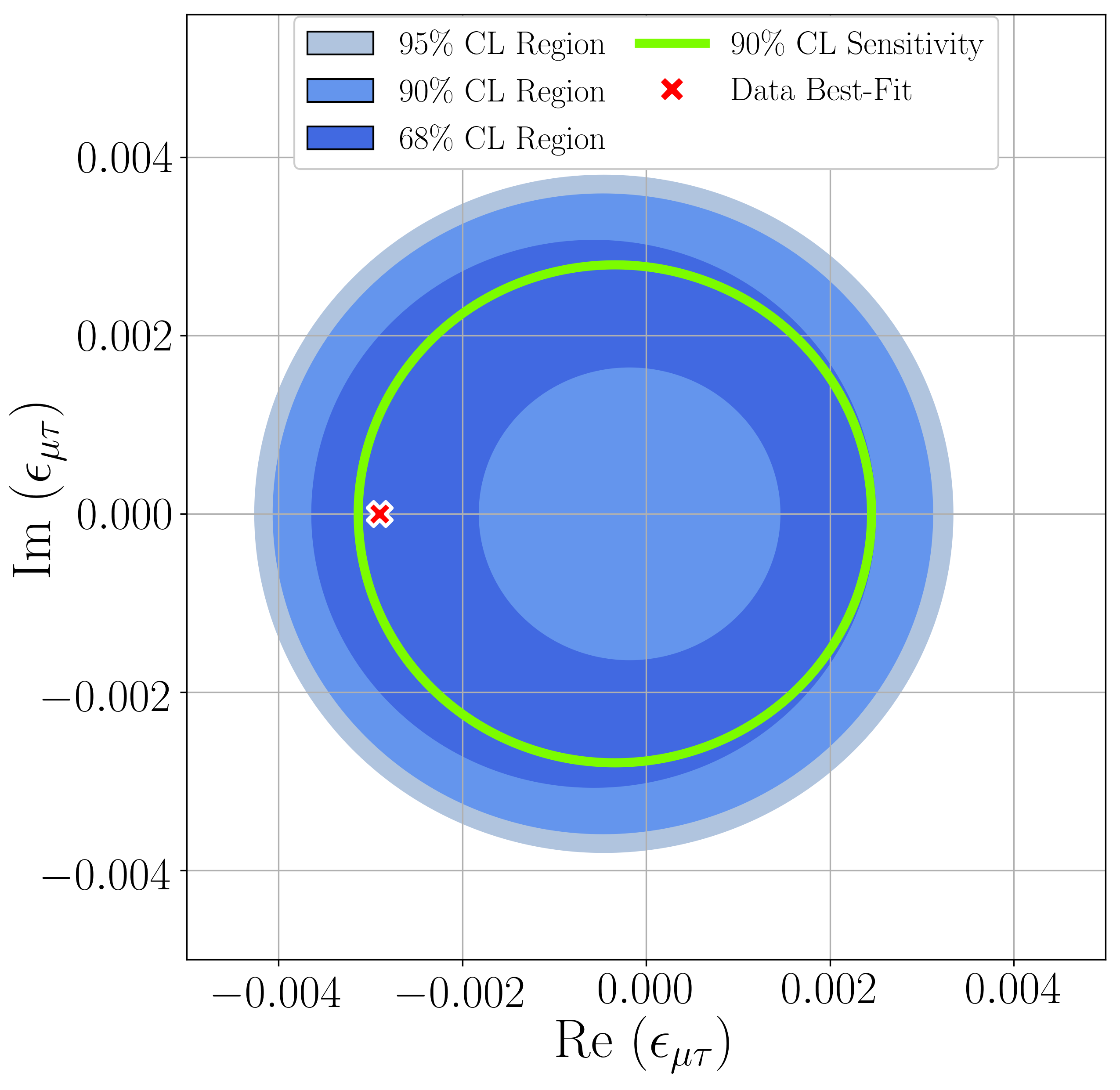}
\caption{\textit{\textbf{Complex result.}}  Confidence level regions for complex $\eut$ in blue-shaded regions, with the analysis 90\% CL sensitivity in green and the red cross marking the data best-fit.}
\label{fig:2d}
\end{figure}

This analysis considers a complex-valued $\eut$ parameter with oscillation probabilities calculated for neutrinos crossing the Earth using the \texttt{nuSQuIDS}~\cite{nusquids,arg2021nusquids} software package.  For illustration, we briefly review the origin of the observed parameter degeneracies using an approximate treatment with small deviations present at the lowest energies, though notably these approximations are not used in the analysis but rather the full 3-neutrino mixing model including matter effects.  From Ref.~\cite{salvado}, the atmospheric neutrino oscillation probability may be approximated for $E_{\nu} > 100$ GeV as

\begin{gather} 
    P(\mnu\rightarrow\nu_\tau) = \left| \sin(2\theta_{23}) \frac{\Delta m^2_{31}}{2E_\nu} + 2V_d \eut  \right|^2 \left( \frac{L}{2}\right)^2~\label{eq:Degen}
\end{gather}

\noindent where $\theta_{23}$ and $\Delta m^2_{31}$ are standard neutrino mixing parameters~\cite{pontecorvo,PMNS}, $E_\nu$ is the neutrino energy, $L$ is the matter baseline, and $V_d$ is the constant potential induced by down quarks (fermion contributions to $\eut$ are normalized to the down quark density, with $N_d \approx 3N_e$ and $N_d \approx N_u$ in Earth~\cite{salvado}).  Changing the mass ordering alters the sign of $\Delta m^2_{23}$, inverting the result across $\epsilon_{\mu\tau}=0$. For complex $\eut$ there is a degeneracy in the complex plane at all energies

\begin{gather}
    P( \eut = a + bi) = P( \eut = a - bi),
\end{gather}

\noindent so all contours, such as in Fig.~\ref{fig:2d}, are symmetric in the imaginary dimension.  Eq.~\ref{eq:Degen} also contains a further degeneracy:  CL boundary contours are circular in the $\epsilon_{\mu\tau}$ complex plane with the center of the circle approaching the origin as $E_\nu\rightarrow\infty$.  The final 2D contour including contributions from all energies is also found to closely resemble a circle with a slight offset from $\text{Re}(\epsilon_{\mu\tau})=\text{Im}(\epsilon_{\mu\tau})=0$.  90\% CL contours from pseudo-experiments adhered sufficiently to a circular form that accurate results could be obtained by testing hypotheses along the real axis only (201 uniformly distributed points in $\text{Re}(\eut) \in [-0.01,0.01] $ with $\text{Im}(\eut) = 0$) and extrapolating the circular contour into the complex plane.  The results were verified from testing 361 uniformly-distributed hypotheses in the full complex space in addition to the aforementioned set, with $\text{Re}(\eut), \text{Im}(\eut)\in [-0.01,0.01] $.  The likelihood threshold for 90\% CL contours was evaluated using the Feldman-Cousins prescription~\cite{feldman} and found to be consistent with Wilks' theorem at one degree of freedom, as expected in the presence of the these degeneracies.

\section{\label{sec:event}Systematic Uncertainties}

Systematic uncertainties are incorporated into the analysis through a collection of nuisance parameters that reweight Monte Carlo (MC) event sets through continuous parameterizations.  The dominant sources of uncertainty derive from the shape and normalization of the atmospheric and astrophysical neutrino fluxes, optical properties of South Pole glacial ice, the local DOM environment, and neutrino interaction cross-sections.  Other sources of systematic uncertainty were investigated and determined to be inconsequential within the overall statistical uncertainty~\cite{meowsprl,meowsprd}. 

\interfootnotelinepenalty=10000
The conventional\footnote{Conventional flux refers to neutrinos produced from $\pi$ and $K$ meson decays in the atmosphere, which is meant to distinguish from the prompt atmospheric flux, referring to neutrinos produced from the decay of atmospheric charmed mesons.} atmospheric $\mnu$ and $\onu$ flux is modeled through pion and kaon decay in the MCEq cascade equation solver~\cite{mceq,mceq2} with the the \texttt{SIBYLL2.3c} hadronic interaction model~\cite{sibyll}.  The spectra of CR primaries relevant to this sample follows an approximate energy dependence of $E^{-2.65}$.  CR spectral index uncertainties are implemented via the nuisance parameter $\Delta\gamma_{\text{conv}}$~\cite{atmg,atmg2,atmg3,atmg4}.   Uncertainties from meson production due to CR-atmosphere and subsequent interactions are accommodated through reweighting fluxes partitioned by incident parent energy and outgoing secondary energy, presented in Ref.~\cite{barr}.  The atmospheric density, relevant to cascade formation, is profiled across zenith through temperature data collected by the AIRS satellite~\cite{airs}.  The corresponding nuisance parameter, \textit{Atm.\ Density}, is introduced through simulated air showers in randomly perturbed density profiles within the provided uncertainty ranges. Kaon energy losses via interaction with atmospheric nuclei are accounted through the total kaon-nucleus cross-section uncertainty~\cite{kaon}.  Uncertainties from charged pion production and interaction are found to be negligible~\cite{meowsprl,meowsprd}.  Lastly, the total conventional atmospheric $\mnu$ and $\onu$ flux has an overall normalization uncertainty~\cite{mceq} quantified by the $\Phi_{\text{conv}}$ parameter.

The astrophysical neutrino spectrum uncertainties are quantified through the normalization ($\Phi_{\text{astro}}$) and spectral index ($\Delta\gamma_{\text{astro}}$) nuisance parameters with correlated Gaussian priors informed by a confidence region encompassing recent IceCube astrophysical flux measurements~\cite{as1,as2,as3,as4,as5,as6}, modeled with a $\mnu:\onu$ ratio of $1:1$ assuming a single-power energy law~\cite{meowsprl,meowsprd}.

The optical properties of the bulk glacial ice result from depth-dependent impurity concentrations.  To minimize the number of relevant parameters and their uncertainties, the absorption and scattering coefficients collected for each 10 m layer are reparameterized into a Fourier series up to a finite cutoff, with modes ordered from the greatest to weakest effects on the propagation of light in the glacial ice.  The SnowStorm software implements an efficient method of sampling the Fourier parameter space by perturbing a single central MC set rather than generating multiple MC sets~\cite{snowstorm}.  Two energy-dependent basis functions are inferred from correlations between perturbed modes, and the amplitudes of these functions ultimately serve as the nuisance parameters for the bulk ice uncertainties.  These nuisance parameters have a bivariate Gaussian prior.

\begin{figure}[t]
\includegraphics[width=\columnwidth]{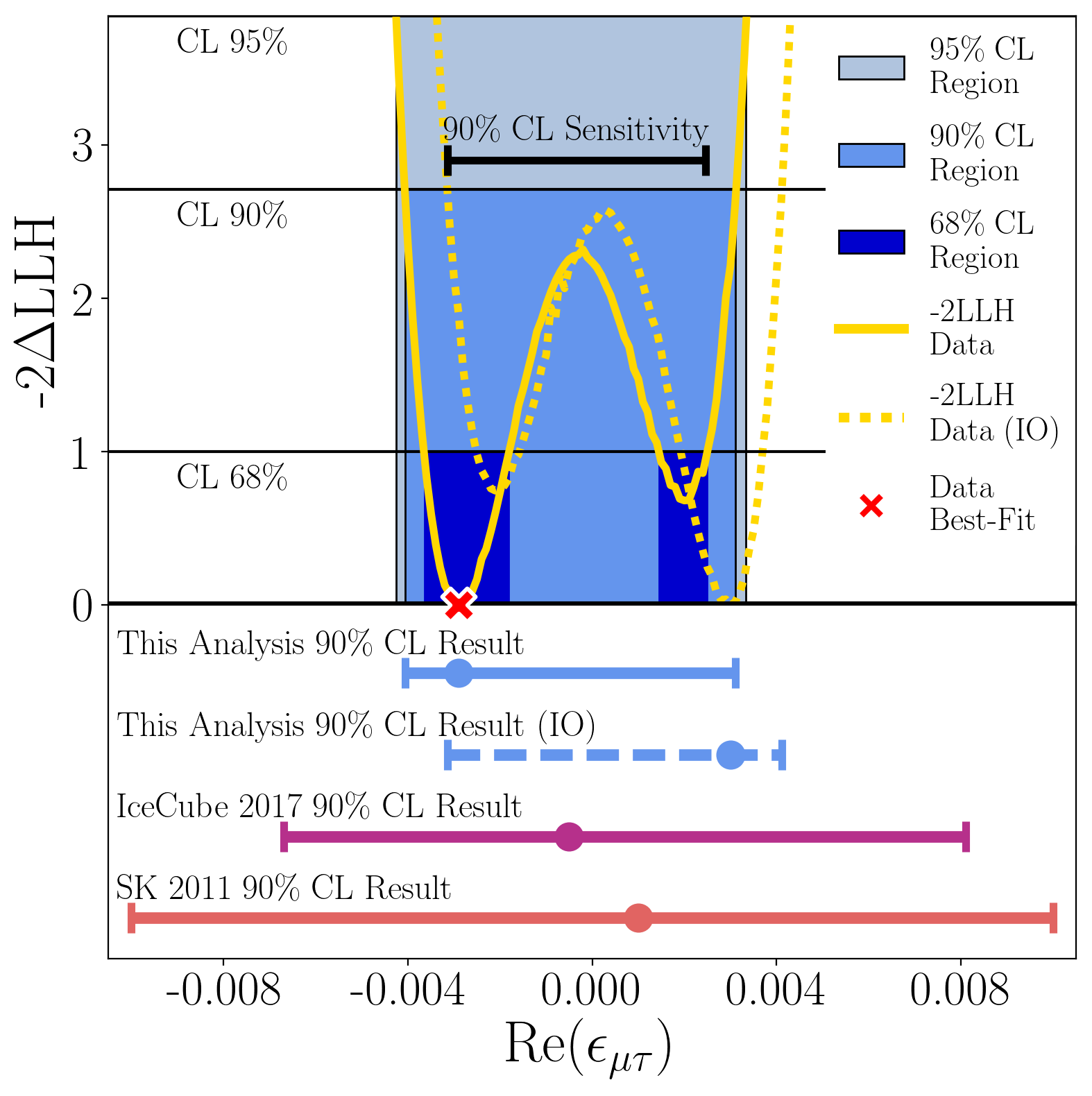}
\centering
\caption{\textit{\textbf{Real-only result.}}  Top: The $-2\Delta LLH$ profile from the fit to data.  Blue-shaded regions correspond to the the CL regions determined from the $-2\Delta LLH$ values.  Bottom: Comparison of the 90\% CL limits from this analysis to IceCube's previous real-only $\eut$ search~\cite{icecube2017} and the Super-Kamiokande experiment's inaugural constraints~\cite{sk2011}.}
\label{fig:1d}
\end{figure}

After deployment, the water in the sensor borehole refreezes with optical impurities inhomogeneously distributed relative to the DOM axis, termed "hole ice"~\cite{bub}.  The consequence of hole ice is the effect on the angular sensitivity in photon detection.  This contribution to the angular efficiency has been modeled empirically with two additional parameters, $p_1$ and $p_2$.  Refs.~\cite{meowsprl,meowsprd} found only one parameter ($p_2$) has a significant contribution to the uncertainty from hole ice such that variations in $p_2$ cover any effects seen in shifts of the negligible parameter ($p_1$).

The uncertainties associated with the effective sensitivity of DOMs to photons after deployment are characterized by the \textit{DOM efficiency} nuisance parameter.  Factors contributing to the efficiency include those internal to the DOM, such as the photocathode efficiency and wavelength acceptance, and factors external to the DOM, including the aforementioned hole ice and sensor cable shadow~\cite{meowsprd,meowsprl}.

The neutrino cross-section determines both the rate of neutrino absorption in Earth~\cite{xs,xs1} and of observable interactions~\cite{xs2,xs3}.  Uncertainties regarding neutrino interactions at the detector were found by Refs.~\cite{jones,ca} to be negligible while the uncertainties of the neutrino cross-sections on in-Earth absorption are parameterized through linearly scaling cross-sections $\sigma_{\mnu}$ and $\sigma_{\onu}$.  The corresponding priors are fixed at the largest uncertainties found within the sample energy range~\cite{xs3}.
 
The impact of the systematic uncertainties was determined by calculating the 90\% CL sensitivity when selected nuisance parameters were fixed while the others were fit freely.  For these tests, the  ``Asimov''~\cite{asimov} sensitivity was employed, following its validation against the true median sensitivity from 1,000 pseudoexperiments.  The most illustrative test fixed categories of parameters organized into three types: hadronic\footnote{W/Y/Z parameters (from Ref.\cite{barr}), atmospheric density, $\Phi_{\text{conv}}$, $\Delta\gamma_{\text{conv}}$}, cosmic\footnote{$\Phi_{\text{astro}}$, $\Delta\gamma_{\text{astro}}$}, and detector\footnote{DOM efficiency, Ice Gradient 0 and Ice Gradient 1 (SnowStorm), $p_2$ (column ice)}. Fixed cosmic nuisance parameters resulted in a $\sim -0.82\%$ relative change in $|\eut|$, while fixed hadronic parameters have a relative change of $\sim -1.63\%$.  Lastly, the largest uncertainty contribution is from the detector parameters, which have a $\sim -9.80\%$ relative change from the central sensitivity radius.  

For a review of the systematic uncertainties treated in this analysis, see Refs.~\cite{meowsprl} and~\cite{meowsprd}.  The prior and posterior widths for the nuisance parameters at the analysis best-fit are listed in the supplementary materials.

\section{\label{sec:results}Results}

\begin{figure}[!ht]
\centering
\includegraphics[width=.48\textwidth]{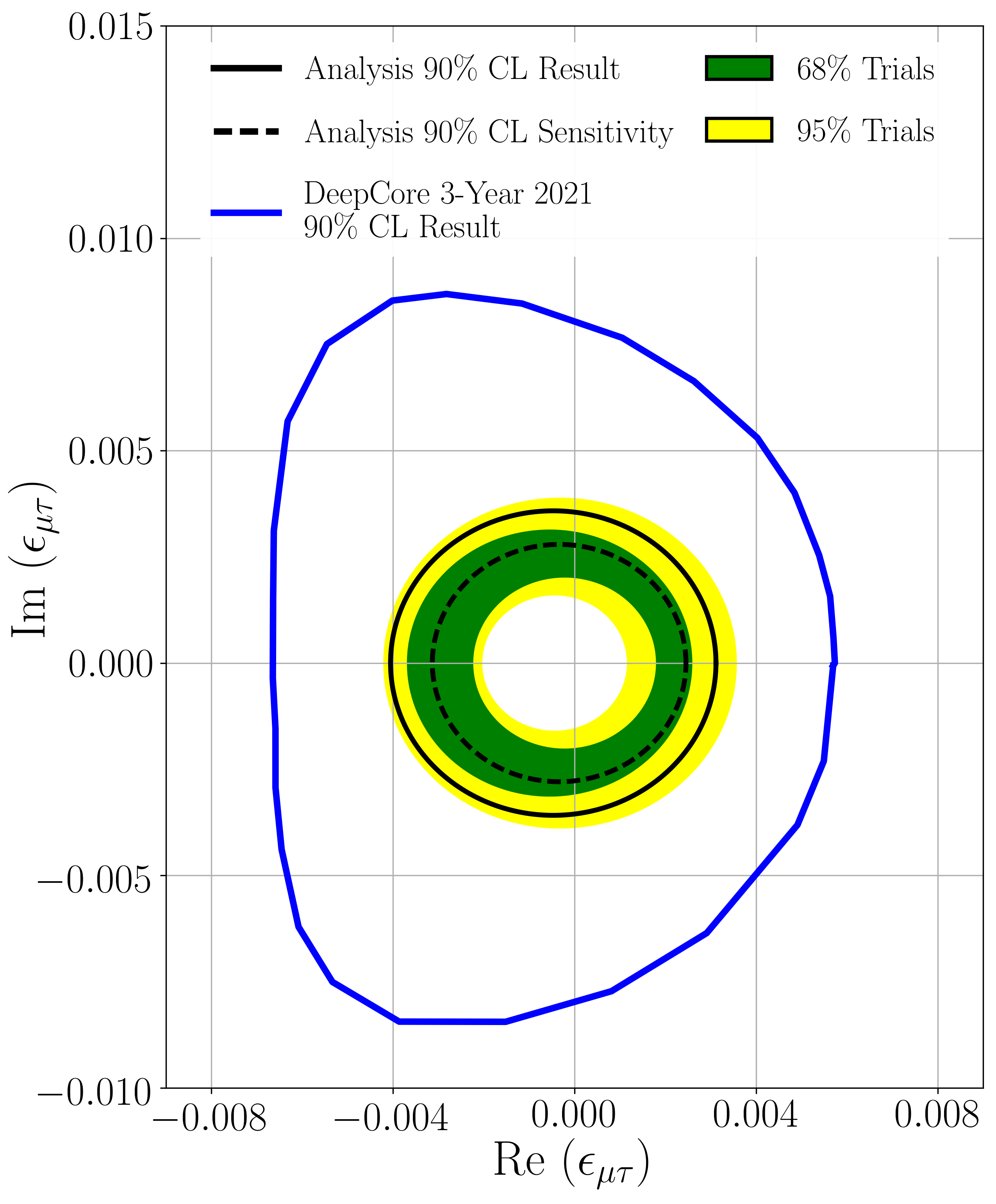}
\caption{\textit{\textbf{Global comparison.}}  Comparison of the analysis 90\% CL sensitivity and result to the DeepCore 3-year, 5.6-100 GeV result~\cite{deepcore2021}.  Green and yellow regions represent 90\% CL sensitivity envelopes of symmetrically-counted 68\% and 95\% (respectively) regions calculated from 1,000 pseudoexperiment trials.}
\label{fig:glo}
\end{figure}

The analysis real-valued best-fit $\textrm{Re}(\eut)=-0.0029$.  The strongest nuisance parameter pull is the cosmic ray spectral index, with a shift of 0.066 $(2.2\sigma)$, while all other systematic uncertainty best-fit values are within $1\sigma$ of their central values.  Fig.~\ref{fig:1d} displays the test statistic profile for the data and the corresponding CL regions in the top panel, followed by a comparison of 90\% CL limits derived from other measurements in the bottom panel.  The analysis limits and sensitivities are a factor of $\sim2$ improvement beyond the leading constraints from Ref.~\cite{icecube2017}. 

In Fig.~\ref{fig:2d} are the CL regions (68\%, 90\%, and 95\%) in complex $\eut$ space.  Fig.~\ref{fig:glo} compares the analysis result and sensitivity to the next-leading complex $\eut$ limits from Ref.~\cite{deepcore2021}, demonstrating an improvement by a factor of $\sim 4$.  The result is found to be consistent with expected experimental sensitivity.  The best-fit \textit{LLH} is found to be -0.68 standard deviations from the distribution mean, which is consistent with no NSI at a p-value of 25.2\% derived from 1000 trial pseudo-experiments. The  best-fit $\eut$ was also consistent with the recovered pseudoexperiment best-fit locations when a non-NSI hypothesis was assumed. 

Compared to initial Re($\eut$) constraints placed by Ref.~\cite{sk2011} and subsequent measurements\footnote{During the revision of this manuscript, Ref.~\cite{ANTARES:2021crm} released Re($\eut$) limits of comparable scale, yet with correlated $\ett$ effects on the results.} such as Refs.~\cite{icecube2017} and~\cite{deepcore2021}, this analysis places the best constraints on Re($\eut$) to-date.  Further, few analyses constrain complex NSI parameters, such as Ref.~\cite{deepcore2021}, and this analysis places the strongest constraints on Im($\eut$) to-date (Fig.~\ref{fig:glo}).

To conclude, 305,735 up-going muon-neutrino tracks from 500 GeV to 9976 GeV detected by the IceCube Neutrino Observatory have been analyzed to search for evidence of $\eut$ NSI.  The best-fit point value is consistent with the no-NSI hypothesis at a p-value of 25.2\%.  The 90\% CL limits on real-only $\eut$ are $-0.0041 < \eut < 0.0031$, representing the strongest constraints on any NSI parameter in any oscillation channel to date.

\section{\label{sec:ack}Acknowledgments}
The IceCube collaboration acknowledges the significant contribution to this manuscript from the University of
Texas at Arlington, Massachusetts Institute of Technology, and Harvard University groups.

We acknowledge the support from the following agencies: USA {\textendash} U.S. National Science Foundation-Office of Polar Programs,
U.S. National Science Foundation-Physics Division,
U.S. National Science Foundation-EPSCoR,
Wisconsin Alumni Research Foundation,
Center for High Throughput Computing (CHTC) at the University of Wisconsin{\textendash}Madison,
Open Science Grid (OSG),
Extreme Science and Engineering Discovery Environment (XSEDE),
Frontera computing project at the Texas Advanced Computing Center,
U.S. Department of Energy-National Energy Research Scientific Computing Center,
Particle astrophysics research computing center at the University of Maryland,
Institute for Cyber-Enabled Research at Michigan State University,
and Astroparticle physics computational facility at Marquette University;
Belgium {\textendash} Funds for Scientific Research (FRS-FNRS and FWO),
FWO Odysseus and Big Science programmes,
and Belgian Federal Science Policy Office (Belspo);
Germany {\textendash} Bundesministerium f{\"u}r Bildung und Forschung (BMBF),
Deutsche Forschungsgemeinschaft (DFG),
Helmholtz Alliance for Astroparticle Physics (HAP),
Initiative and Networking Fund of the Helmholtz Association,
Deutsches Elektronen Synchrotron (DESY),
and High Performance Computing cluster of the RWTH Aachen;
Sweden {\textendash} Swedish Research Council,
Swedish Polar Research Secretariat,
Swedish National Infrastructure for Computing (SNIC),
and Knut and Alice Wallenberg Foundation;
Australia {\textendash} Australian Research Council;
Canada {\textendash} Natural Sciences and Engineering Research Council of Canada,
Calcul Qu{\'e}bec, Compute Ontario, Canada Foundation for Innovation, WestGrid, and Compute Canada;
Denmark {\textendash} Villum Fonden and Carlsberg Foundation;
New Zealand {\textendash} Marsden Fund;
Japan {\textendash} Japan Society for Promotion of Science (JSPS)
and Institute for Global Prominent Research (IGPR) of Chiba University;
Korea {\textendash} National Research Foundation of Korea (NRF);
Switzerland {\textendash} Swiss National Science Foundation (SNSF);
United Kingdom {\textendash} Department of Physics, University of Oxford.

\appendix*

\section{\label{sec:system}Appendix: Systematic Uncertainty Prior and Posterior Widths}
\begin{table}[h!]
\caption{\label{tab:sys} The analysis nuisance parameters with corresponding central values and $1\sigma$ widths of the prior and posterior distributions.  Choices of priors are reported in Refs. \cite{meowsprd,meowsprl}.  The $\pm1\sigma$ posterior uncertainties are calculated from distributions obtained from 1000 pseudoexperiments simulated at the model best-fit.  Listed parameter constraints are Gaussian distributed except for those with an asterisk, denoting a bivariate Gaussian distribution.}
\setlength{\tabcolsep}{15pt}
\begin{ruledtabular}
\begin{tabular}{lcr}
\textbf{Parameter} & \textbf{Constraint} & \textbf{Best-Fit} $\mathbf{\pm 1 \sigma}$ \\
\end{tabular}
\end{ruledtabular}
\newline
\vspace*{1mm}
\textbf{Detector Parameters}
\vspace*{1mm}
\begin{ruledtabular}
\begin{tabular}{lcr}
DOM Efficiency & $0.97 \pm 0.10$ & $0.965 \pm 0.006$\\
\hline
Ice Gradient 0   & $0.0 \pm 1.0* $& $0.07 \pm 0.23$\\
\hline
Ice Gradient 1 &  $0.0 \pm 1.0* $ & $0.80 \pm 0.41$\\
\hline
Hole Ice ($p_2$)  &$ -1.0 \pm 10.0 $ & $-3.20 \pm 0.46$\\
\end{tabular}
\end{ruledtabular}
\vspace{2mm}
\textbf{Atmospheric. Flux Parameters}
\vspace{2mm}
\begin{ruledtabular}
\begin{tabular}{lcr}
Norm. $(\Phi_{\text{conv}})$ & $1.0 \pm 0.4$ & $1.11\pm$0.04 \\
\hline
Shift ($\Delta \gamma_{\text{conv}}$) & $ 0.00 \pm 0.03 $ & $0.066 \pm 0.008$ \\
\hline
Atm. Den. &  $0.0 \pm 1.0$  &  $-0.08\pm0.41$\\
\hline
Atm. WM  & $0.00 \pm 0.40$ &  $ -0.02 \pm 0.05$\\
\hline
Atm. WP  & $0.00 \pm 0.40$ &  $0.01 \pm 0.04$\\
\hline
Atm. YM  & $0.00 \pm 0.30$ &  $-0.06 \pm 0.06$\\
\hline
Atm. YP  & $0.00 \pm 0.30$ &  $-0.06 \pm 0.06$\\
\hline
Atm. ZM  & $0.00 \pm 0.12$ &  $-0.01 \pm 0.01$\\
\hline
Atm. ZP  & $0.00 \pm 0.12$ &  $-0.02 \pm 0.01$\\
\hline
E. Loss $\sigma_{\text{KA}}$  & $0.0 \pm 1.0$ &  $-0.09\pm 0.15$\\
\end{tabular}
\vspace{2mm}
\textbf{Astrophysical Flux Parameters}
\vspace{2mm}
\begin{tabular}{lcr}
Norm. ($\Phi_{\text{astro}}$) & $0.00 \pm 0.36*$ &  $0.86 \pm 0.13$\\
\hline
Shift ($\Delta \gamma_{\text{astro}}$)  & $0.00 \pm 0.36*$ &  $0.03 \pm 0.14 $\\
\end{tabular}
\vspace{2mm}
\textbf{Neutrino Cross Sections}
\vspace{2mm}
\begin{tabular}{lcr}
Cross-Sec $\sigma_{\nu_\mu}$ & $1.00 \pm 0.03$ &  $1.001 \pm 0.002$\\
\hline
Cross-Sec $\sigma_{\overline{\nu}_\mu}$  & $1.000 \pm 0.075$ &  $1.001 \pm 0.005$\\
\end{tabular}
\end{ruledtabular}
\end{table}

\nocite{*}

\providecommand{\noopsort}[1]{}\providecommand{\singleletter}[1]{#1}%

\end{document}